\documentclass[a4paper]{spie}  
\usepackage[]{graphicx}
\usepackage{textcomp}
\usepackage{amsmath,amsfonts,amssymb}
\usepackage{multirow}

\title{Progress on the Astrometric Gravitation Probe design} 

\author{M. Gai\supit{a}, A. Vecchiato\supit{a}, A. Riva\supit{a}, M.G. Lattanzi\supit{a}, 
	F. Landini\supit{a}, B. Bucciarelli\supit{a}, D. Busonero\supit{a}, M. Crosta\supit{a}, 
	S. Liao\supit{b}, H. Luo\supit{b}, G. Mana\supit{c}, M. Pisani\supit{c}, Z. Qi\supit{b}, 
	C.P. Sasso\supit{c}, Z. Tang\supit{b}, Y. Yu\supit{b}
\skiplinehalf
\supit{a}INAF - Osserv. Astrofisico di Torino, V. Osservatorio, 20, 10025 Pino Torinese (TO), Italy \\
\supit{b}Shanghai Astronomical Observatory, CAS, 80 Nandan Rd, Shanghai 200030, China\\
\supit{c}INRiM - Ist. Nazionale di Ricerca Metrologica, Str. delle Cacce, 91, 10135 Torino (TO), Italy \\
}
\authorinfo{Further author information: E-mail: mario.gai@inaf.it, Telephone: +39 011 8101 943, www.oato.inaf.it}

\begin{document} 
\maketitle 

\begin{abstract}

The Astrometric Gravitation Probe mission is a modern version of the 1919 Dyson-Eddington-Davidson experiment, based on a space-borne telescope with a permanent built-in eclipse, provided by a coronagraphic system. The expected improvement on experimental bounds to General Relativity and competing gravitation theories is by at least two orders of magnitude. The measurement principle is reviewed, in particular the principle of Fizeau-like combination of a set of individual inverted coronagraphs simultaneously feeding a common high resolution telescope. Also, the payload has a dual field of view property, in order to support simultaneous observations of stellar fields either very close, or far away, from the Sun, i.e. fields affected by either high or low light bending. We discuss a set of solutions introduced in the optical design to improve on technical feasibility and robustness of the optical performance against perturbations, in particular induced by manufacturing and alignment tolerances, and launch stresses. 

\end{abstract}



\section{Introduction}
\label{sec:introduction}  

The Astrometric Gravitation Probe (AGP) concept defines a space mission aimed at Fundamental Physics testing, and in particular General Relativity (GR), to unprecedented levels. Astrometry is intrinsically suited for precision testing of GR that exploits the so-called light deflection effect, as well as that involving the motion of the Solar System bodies (see e.g.\ Crosta\cite{2019NCimR..42..443C} and references therein).

In the Solar System environment it is customary and a convenient practice to model such phenomena in the framework of the so-called ``Parametrized Post-Newtonian (PPN) formalism'',\cite{Will2006} which relies on 10 parameters to model and quantify many possible deviations to the predictions of GR at the Post-Newtonian order. 
The light deflection effect is related to the PPN $\gamma$ parameter ($=1$ in GR), whose most precise estimation is currently based on time delay measurements on artificial probes, e.g. Cassini,\cite{Bertotti2003} reaching a precision $\sigma_\gamma/\gamma\simeq 10^{-5}$.

It may be noted that general purpose experiments, e.g.\ Gaia,\cite{2003AAp...399..337V,2010IAUS..261..315H} although in principle able to improve on current experimental achievements up to $\sim10^{-6}$, may fall short on this subject due e.g.\ to measurement correlation.\cite{Butkevich2021inPrep}
Such level of precision barely enters in the range $10^{-5}<1-\gamma<5\cdot10^{-8}$, mentioned in Turyshev et al.\cite{2009ExA....27...27T} and in references therein, 
at which deviations are expected, either in GR or in other relevant parts of our basic understanding of Nature's laws, e.g.\ particle physics, if an overarching Unified Theory is to be established. A different kind of approach is thus needed for an in-depth probing of this range.

Here we wish to show one of the possible approaches, and to give an estimation of its potential performance. The issue of high accuracy astrometry is being investigated by the AGP team also in the context of the ASTRA (Astrometric Science and Technology Roadmap for Astrophysics) project\cite{Gai2020ASTRA}. In particular, a telescope especially suited to astrometry over a large annular field of view, thanks to uniform optical response, is appealing, and the concept is being investigated in detail\cite{Riva2020RAFTER}. 

\subsection{Order-of-magnitude considerations}
\label{sec:order-of-mag}  

\begin{figure}
  \begin{center}
    \includegraphics[width=0.95\textwidth]{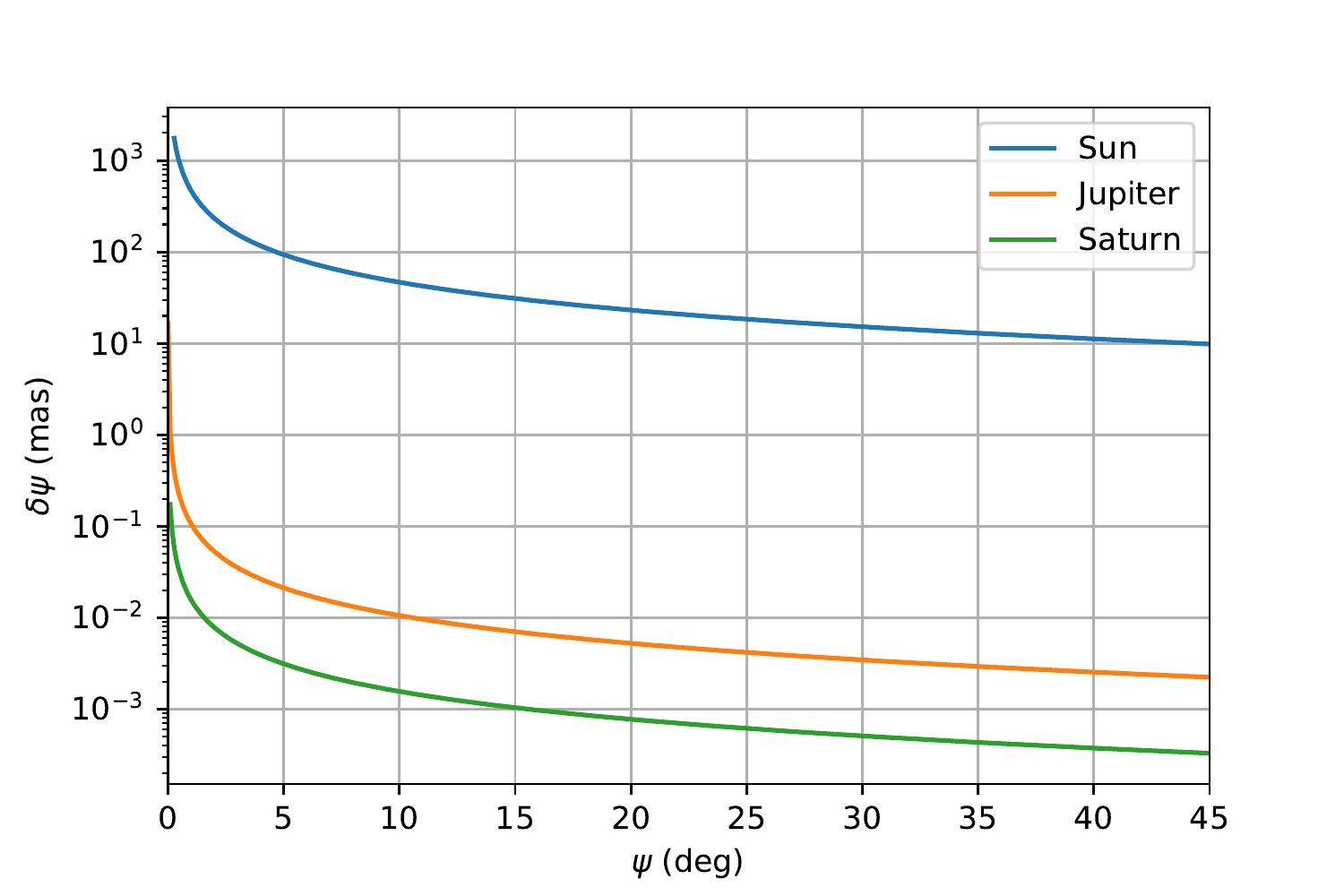}
    \caption{\label{fig:deflSS} Light deflection around the main bodies of the Solar System from an observer at the Earth.}
  \end{center}
\end{figure}

Following Misner et al.\cite{Misner1973}, the amount of light deflection, namely the difference $\delta\psi=\psi_{\mathrm{o}}-\psi$ between the observed stellar direction $\psi_{\mathrm{o}}$, deflected by the mass of the Sun $M_{\odot}$ and seen from the Earth, and the undeflected one $\psi$, can be written as function of the PPN-$\gamma$ parameter 
as
\begin{equation}
   \delta\psi=
      \frac{\left(1+\gamma\right)GM_{\odot}}{c^{2}r_{\oplus}}
      \left(\frac{1+\cos\psi}{1-\cos\psi}\right)^{1/2},
\label{eq:deflMisner}
\end{equation}
where $r_{\oplus}$ is the distance of the Earth from the Sun. 
This formula easily shows that the direct astrometric effect of light deflection, namely the displacement from their apparent position of objects observed close to other sources of gravity, is proportional to the mass of the deflecting object, and it is also larger for smaller elongations $\psi$. So, substituting the values of each specific body of the Solar System, we can obtain a figure similar to Fig.~\ref{fig:deflSS} in which the effects of the different Solar System bodies are plotted at different elongations.

It is easy to understand that the weak point of the Gaia approach with respect to determination of the PPN $\gamma$ parameter is the large angular distance of the satellite's line of sight from the main source of gravity of the Solar System, 
i.e.\ the Sun, and that a mission designed to get close to the solar edge would be much more fit to this purpose. 
We thus want to find a more quantitative way to predict the possible performance of such a mission.

Assuming that the uncertainty on the estimation of the deflection (i.e.~$\sigma_{\delta\psi}$) depends only on $\sigma_{\gamma}$ and $\sigma_{\psi}$, namely the uncertainty on $\gamma$ and the angular measurement accuracy respectively, it is
\begin{equation}
  \sigma_{\delta\psi}=
     \sqrt{\left(\frac{\partial\left(\delta\psi\right)}{\partial\psi}\right)^{2}\sigma_{\psi}^{2}+
           \left(\frac{\partial\left(\delta\psi\right)}{\partial\gamma}\right)^{2}\sigma_{\gamma}^{2}}.
\label{eq:sigma_dpsi_1}
\end{equation}

After straightforward calculations one finds
\begin{align}
  \frac{\partial\left(\delta\psi\right)}{\partial\psi} &
     =\delta\psi\frac{\left(-\sin\psi\right)}{\left(1-\cos\psi\right)\left(1+\cos\psi\right)} \label{eq:ddeltapsidpsi} \\ 
  \frac{\partial\left(\delta\psi\right)}{\partial\gamma} &
     =\left(1+\gamma\right)^{-1}\delta\psi,\label{eq:ddeltapsidgamma}
\end{align}
which, by substitution in Eq.~(\ref{eq:sigma_dpsi_1}), give
\begin{equation}
  \left(\frac{\sigma_{\delta\psi}}{\delta\psi}\right)^{2}=
    \left[\frac{\left(-\sin\psi\right)}{\left(1-\cos\psi\right)\left(1+\cos\psi\right)}\right]^{2}\sigma_{\psi}^{2}+
    \left(1+\gamma\right)^{-2}\sigma_{\gamma}^{2}
\label{eq:rel_error_1}
\end{equation}

This expression can be converted into a useful relation between $\sigma_{\gamma}$ and the ratio $\sigma_{\psi}/\left|\delta\psi\right|$ by considering that $\delta\psi=\psi_{\mathrm{o}}-\psi$. In fact, one can assume either that the deflection is obtained by two different angular measurements or just by one that has to be compared with some catalog values having negligible errors, giving $\sigma_{\delta\psi}=\sqrt{2}\sigma_{\psi}$ or $\sigma_{\delta\psi}=\sigma_{\psi}$ respectively. Substituting this relation into Eq.~(\ref{eq:rel_error_1}), and isolating $\sigma_{\gamma}^{2}$, we obtain
\begin{equation}
   \sigma_{\gamma}^{2} = 2\left(1+\gamma\right)^{2}
     \left\{ 1-\frac{1}{2}\left[\frac{\delta\psi\,\sin\psi}{\left(1-\cos\psi\right)\left(1+\cos\psi\right)}\right]^{2}\right\} \frac{\sigma_{\psi}^{2}}{\left(\delta\psi\right)^{2}},
\label{eq:sigma_gamma_sq_1}
\end{equation}
where obviously the factor 2 disappears in case we assume $\sigma_{\delta\psi}=\sigma_{\psi}$.

For small angles $\psi\ll1$ one can exploit the Taylor expansions of $\sin\psi$ and $\cos\psi$ to transform Eq.~\eqref{eq:sigma_gamma_sq_1} into the simpler approximate relation
\begin{equation}
  \sigma_{\gamma}\simeq\sqrt{2}\left(1+\gamma\right)\sqrt{1-\frac{1}{2}\left(\frac{\delta\psi}{\psi}\right)^{2}}\frac{\sigma_{\psi}}{\left|\delta\psi\right|},
\label{eq:sigma_gamma_1}
\end{equation}
and since we are working in the weak gravity regime, we can safely assume $\delta\psi\ll\psi$, which implies
\begin{equation}
  \sqrt{1-\frac{1}{2}\left(\frac{\delta\psi}{\psi}\right)^{2}}=
    1-\frac{1}{2}\left(\frac{\delta\psi}{\psi}\right)^2+\mathcal{O}\left(\left(\frac{\delta\psi}{\psi}\right)^{4}\right),
\label{eq:approx_dpsioverpsi}
\end{equation}
so that
\begin{equation}
\sigma_{\gamma} \simeq2
  \sqrt{2}\frac{\sigma_{\psi}}{\left|\delta\psi\right|},
\label{eq:sigma_gamma_3}
\end{equation}
where in Eq.~(\ref{eq:sigma_gamma_3}) we have used the approximation $\gamma\simeq1$. Again, in case we assume $\sigma_{\delta\psi}=\sigma_{\psi}$ the $\sqrt{2}$ has to be dropped from this formula.

Eq.~\eqref{eq:sigma_gamma_3} is a very simple formula that can be used to estimate the accuracy on the PPN-$\gamma$ parameter attainable by an instrument that can measure angles with an accuracy of $\sigma_\psi$ when the light deflection effect amounts to $\delta\psi$. For example, Gaia can reach a single-measurement accuracy of $\sim10^{-1}\,mas$ for $G\simeq13$ objects, which implies a $10^{-1}\lesssim\sigma_\gamma\lesssim\cdot10^{-2}$ in the regions observed by the satellite, where the elongation from the Sun is some tens of degrees and consequently $\delta\psi\sim2\,mas$. It is therefore easy to estimate that this mission needs to cumulate roughly $\sim10^{9}$ observations to obtain the claimed $\sigma_\gamma\sim10^{-6}$ accuracy.

\section{Optical Design and Operation}
\label{sec:OptDesign}  
The proposed operating mode is based on observation of sources over a ring around the Sun, rejecting the latter thanks to a coronagraphic system described in the literature\cite{AGP_Landini16}. We set the annular field of view with a $\sim 1^\circ$ radius around the optical axis, pointed close to the Sun center. Following the Sun throughout the year, all objects in a strip centered on the Ecliptic are observed, all contributing similarly to the measurement of deflection, as each of them is observed at a narrow angle ($\sim 1^\circ$) from the deflecting mass, and therefore subject to light bending by about $460\,mas$, with circular symmetry orientation. 

The specifications for an annular field in AGP have been used for development of the RAFTER telescope configuration, in subsequent stages to allow for verification. 

\begin{figure}
  \begin{center}
    \includegraphics[width=0.85\textwidth]{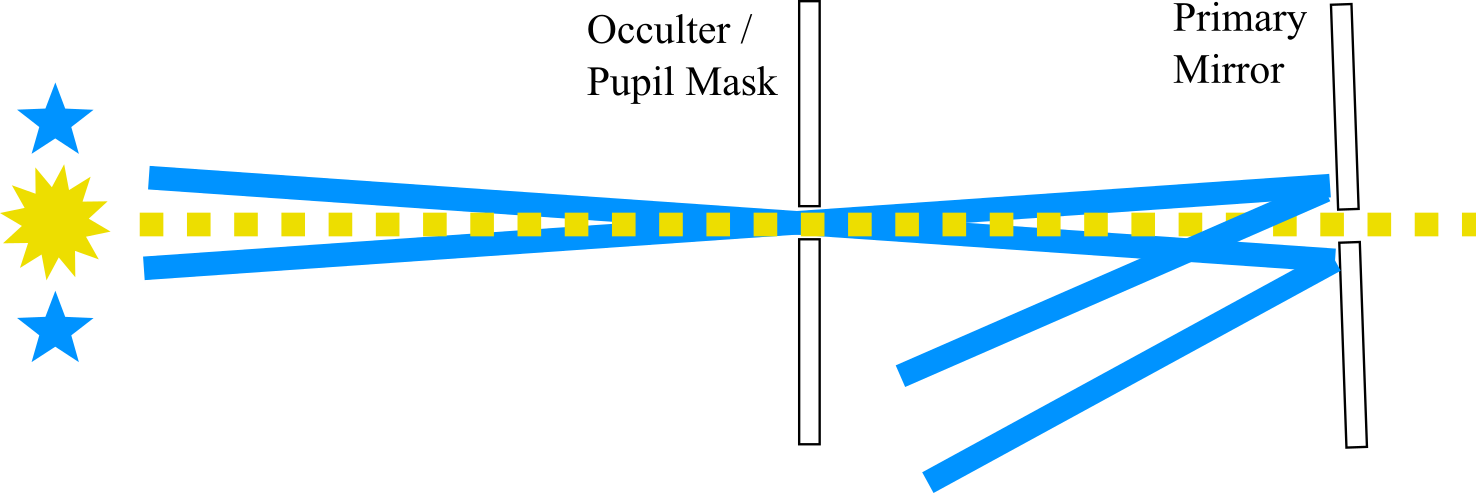}
    \caption{\label{fig:InvOcc} Principle of the inverted occulter}
  \end{center}
\end{figure}

\subsection{The Inverted Occulter Principle}
\label{sec:InvertOccult}
The inverted occulter principle is mutuated by the innovative occulting design of the Solar Orbiter/Metis solar coronagraph \cite{Antonucci_2020, Fineschi_2020}. The main goal of a solar coronagraph is to observe the solar corona, which is an optically thin mean, very close in angle to the much brighter (more than 6 orders of magnitude) photosphere. This generates the extreme need for stray light reduction and control. In the classical external occultation scheme for coronagraphs, an opaque diaphragm is blocking the solar disk light. The inverted scheme foresees the inversion of the entrance aperture and the opaque diaphragm. The solution, while keeping a low stray light level, allows to maintain a compact dimension of the instrument, which is always a plus in space instrumentation. Stray light is mainly generated by the diffraction of the photospheric light from the edges of the surface that is directly hit by the solar disk light. 
In the case of Metis, the main stray light source is the external aperture, that is directly facing the Sun. The inverted occulter principle in the case of AGP allows to dump out of the telescope the photospheric light, by letting it pass through the apertures (see Fig. \ref{fig:InvOcc}). Notably, diffraction from the circular entrance apertures is not totally removed, and need a careful estimate in order to properly size the apertures dimension, or -if needed- adopt further occulting optimization techniques\cite{Metis_Landini2016}. 

The occultation concept for AGP is schematically shown in Fig.\,\ref{fig:InvOcc}, in which the beam from the Sun, marked by the yellow dashed line, gets into the pupil mask (PM) aperture and out of the corresponding aperture on the primary mirror (M1), whereas the beams from field stars at an adequate angular distance (light blue solid lines) are collected by the mirror surface and, following the telescope optics, imaged onto the detector. The unvignetted beams are shown in figure for two sources, respectively above and below the Sun in the drawing plane; actually, circular symmetry ensures that the whole annular field (at $\sim 1^\circ$ from the Sun) is observed simultaneously, illuminating an annular surface on M1 around the aperture. Exit apertures are larger than entrance apertures due to the finite angular size of the Sun (plus margins). 

\begin{figure}
  \begin{center}
    \includegraphics[width=0.45\textwidth]{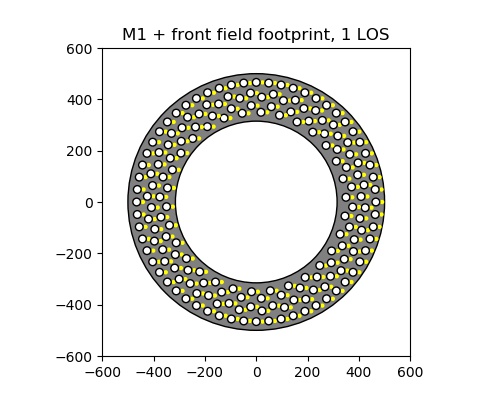}
    \includegraphics[width=0.45\textwidth]{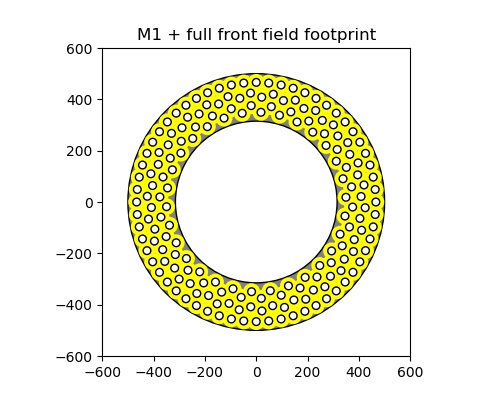}
    \caption{\label{fig:Footprint} Footprint on the primary mirror from one LOS (left) and 
    from the whole annular field (right) }
  \end{center}
\end{figure}

The aperture pair (respectively on PM and M1) can be replicated several times onto the geometry of a larger diameter telescope. All apertures have the same geometric constraints and manufacturing tolerances, and the two sets are aligned simultaneously, since each aperture is in a fixed position with respect to all others by construction. 

\subsection{The Fizeau Interferometer Principle}
\label{sec:Fizeau}
The beam footprint from one line-of-sight (LOS) is shown in Fig.\,\ref{fig:Footprint} (left), as a set of yellow spots on one side of each aperture on M1; the beam footprint associated to the whole annular field of view makes a better usage of the M1 surface, as shown on the right panel. The adopted geometry corresponds to the final optical design ($1\,m$ diameter) described in the following. 

The collecting area for each LOS is small, as it is just the entrance aperture size replicated by the number of apertures; besides, the resolution remains that of the underlying telescope, used as a Fizeau (stellar) interferometer. In particular, the imaging quality of the telescope corresponds to the cophasing performance of the interferometer, associated to the fringe contrast. 

\begin{figure}
  \begin{center}
    \includegraphics[width=0.45\textwidth]{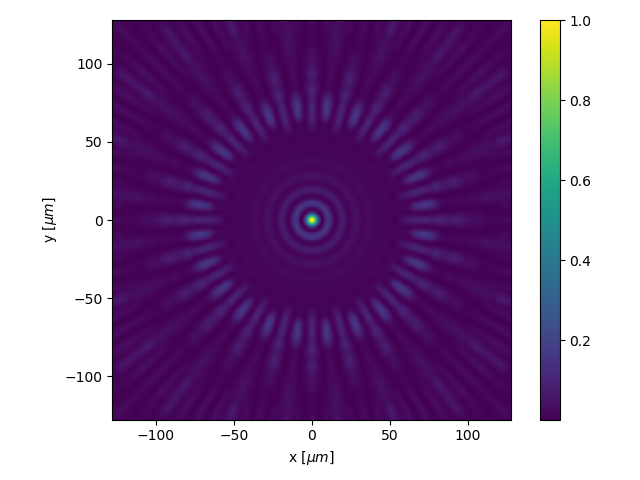}
    \includegraphics[width=0.45\textwidth]{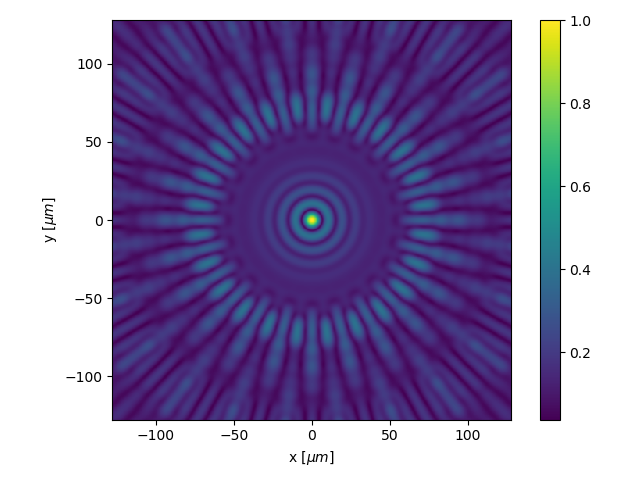}
    \caption{\label{fig:PSFpoly} Polychromatic PSF in linear (left) and 
    square root (right) scale }
  \end{center}
\end{figure}

The ideal polychromatic PSF provided by the selected geometric arrangement is shown in Fig.\,\ref{fig:PSFpoly}, respectively in linear scale (left) and in square root scale (right), for a near-solar spectral type star, using a $20\%$ bandwidth. The coherent set of apertures retains the underlying telescope resolution, providing a bright central spot and a few diffraction rings of a $1\,m$ diameter instrument; significant modulation is retained over the large region associated to the PSF of the much smaller entrance aperture. 

\subsection{Telescope Optical Design Development}
\label{sec:TelOptDes}
The optical design rationale and concept is described in another contribution to these Proceedings.\cite{Riva2020RAFTER} We recall that the starting point is a Three-Mirror Anastigmat (TMA)\cite{Korsch77,Korsch80}, further developed under the prescription of achieving
\begin{enumerate}
    \item a diffraction limited annular field of view with $\sim 1^\circ$ radius;
    \item a compact configuration.
\end{enumerate}

\begin{figure}
  \begin{center}
    \includegraphics[width=0.85\textwidth]{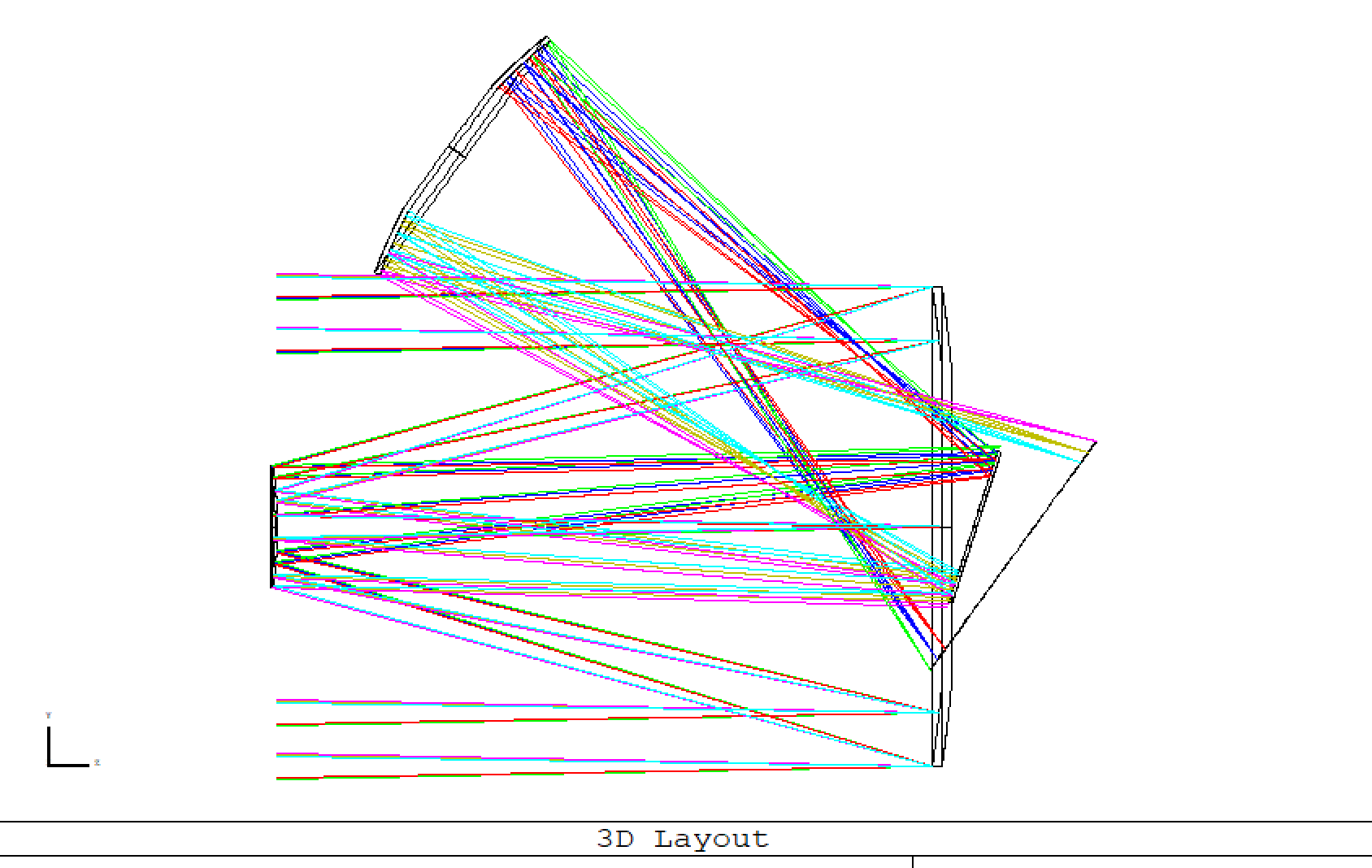}
    \includegraphics[width=0.85\textwidth]{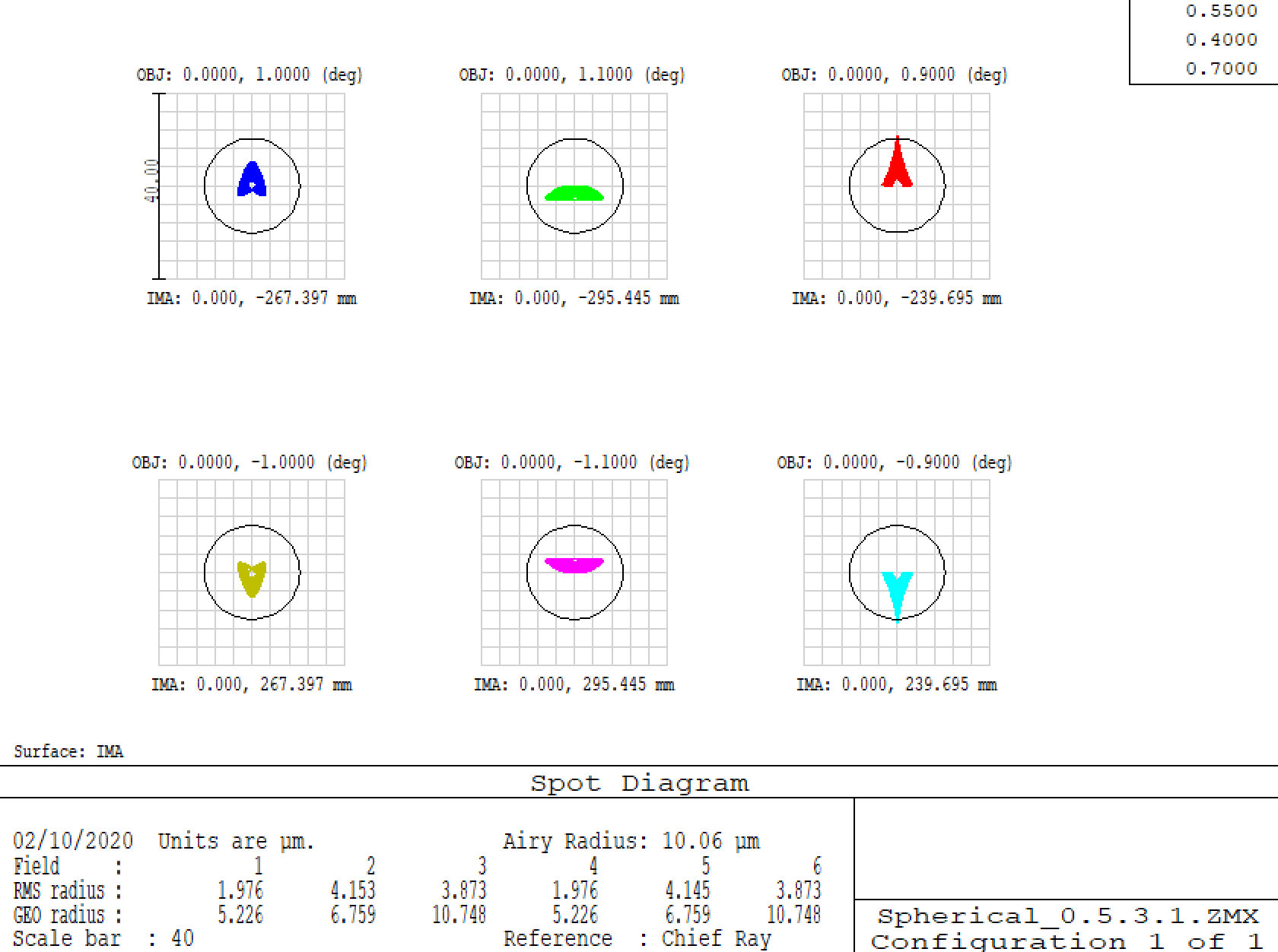}
    \caption{\label{fig:cfg070} Optical layout of 70 cm configuration (top) and 
    spot diagram (bottom) }
  \end{center}
\end{figure}

\begin{figure}
  \begin{center}
    \includegraphics[width=0.85\textwidth]{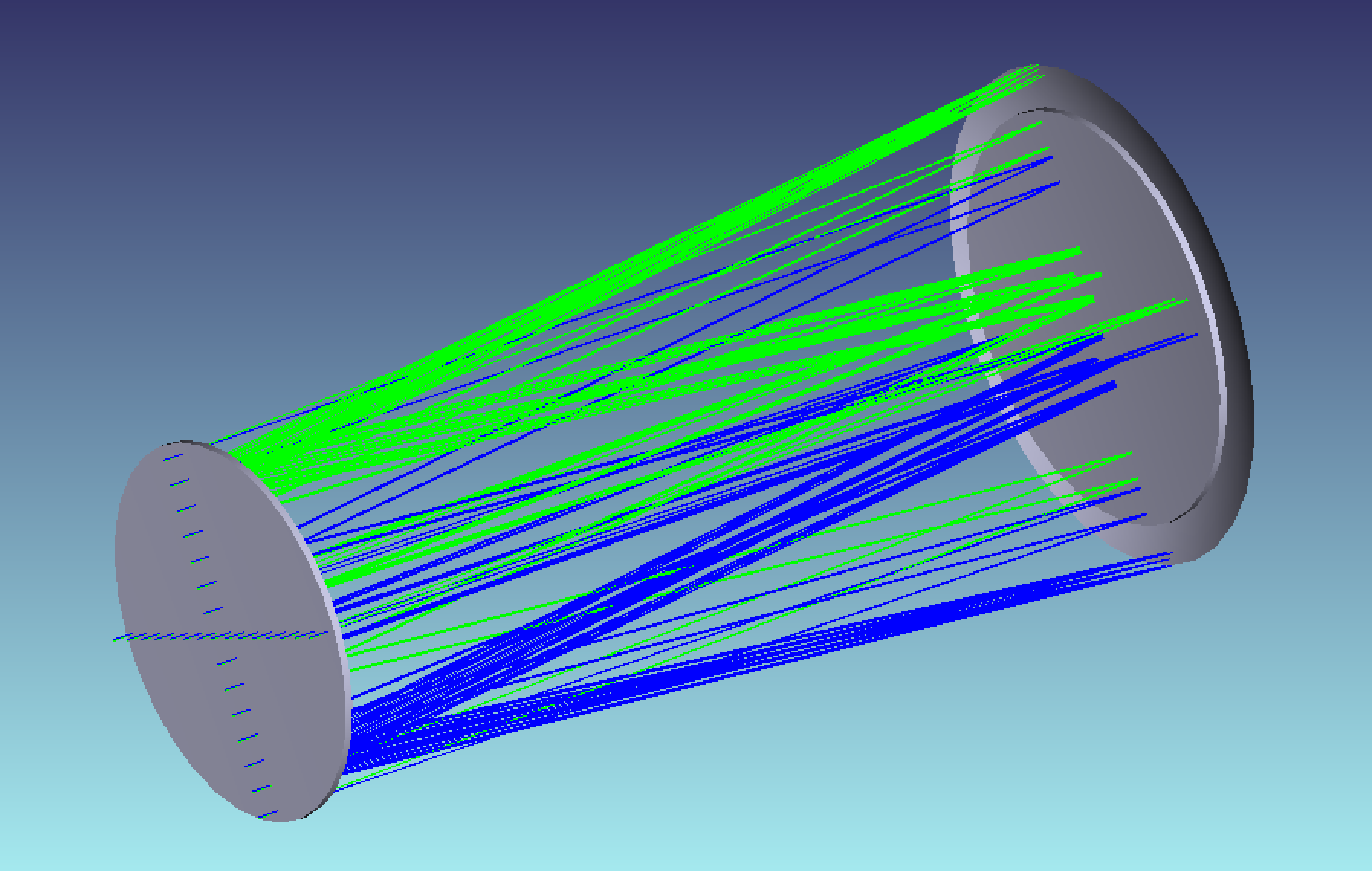}
    \includegraphics[width=0.85\textwidth]{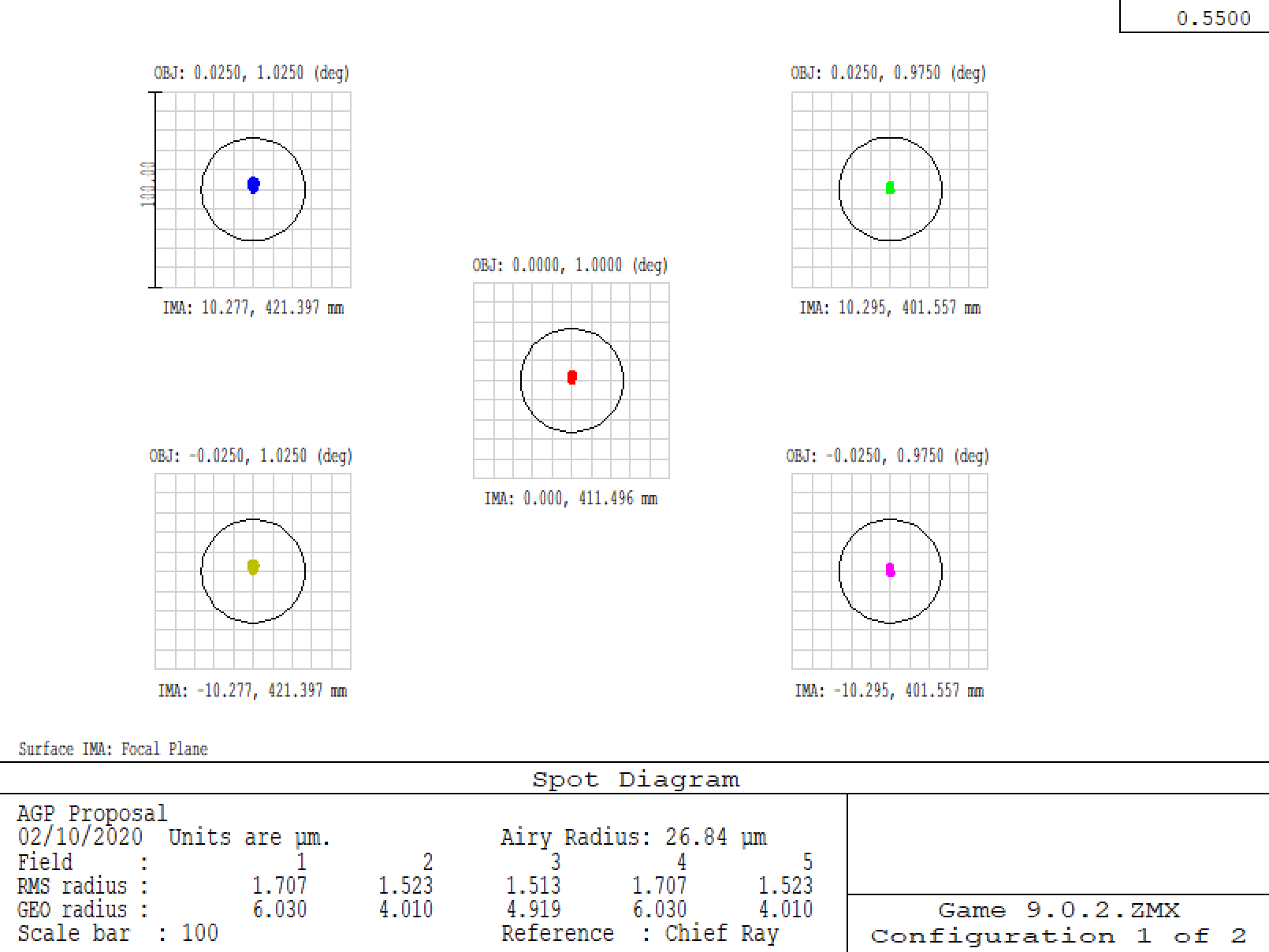}
    \caption{\label{fig:cfg100} Optical layout of 100 cm configuration (top) and 
    spot diagram (bottom) }
  \end{center}
\end{figure}

The first step was to modify the two-axis TMA geometry in order to reduce the volume occupation. The intermediate result is shown in Fig.\,\ref{fig:cfg070} (top), in which the folding mirror (FM) after the secondary mirror (M2) deviates the optical axis by an angle of about $30^\circ$, placing the tertiary mirror (M3) close to the input beam. 
This configuration has primary mirror (M1)  diameter $D_1=0.70\,m$ and primary to secondary distance $d_{12}=2\,m$, thus fitting an envelope of $\sim 2.2\,m$. 
The optics layout is shown at the top of Fig.\,\ref{fig:cfg070}, with the corresponding spot diagram shown below. The field considered positions are the central position and edges of an annular region between $0^\circ.9$ and $1^\circ.1$, enclosing a total area of $1.257$ square degrees. This area is diffraction limited, since the polychromatic ray distribution lays well within the Airy circle. The symmetry of the optical response is evidenced by the spot diagrams, mirror images of each other in opposite radial positions. 

The optical design has then been further pushed to a larger primary mirror diameter $D_1=1\,m$, aiming at a higher imaging resolution and a better astrometric performance, and shorter primary to secondary distance $d_{12}=1.40\,m$, thus fitting a more compact envelope of $\sim 1.8\,m$. 
The system is made even more compact by folding back the second axis onto the original optical axis, so that M3 is allocated close to M2 (and supported by the same spiders). A working configuration has been found, without vignetting. The larger central obscuration from the M3/M2 assembly is considered an acceptable penalty for the very compact geometry, which also retrieves overall axial symmetry. 

The 3D layout of the optical system is shown in the top panel of Fig.\,\ref{fig:cfg100}, whereas the monochromatic spot diagrams at $\lambda = 550\,nm$ are shown in the right panel. The area considered is a $3 \times 3\, arcmin$ region around the central ring position at radius $1^\circ$, corresponding to one chip of the detection system; the actual diffraction limited area remains the same as for the previous configuration, i.e.\ the $1.257$ square degrees focal plane between $0.9^\circ$ and $1.1^\circ$. 

The concept of an annular field telescope is investigated in a dedicated contribution to these Proceedings,\cite{Riva2020RAFTER} based on the above $D_1=1\,m$ configuration. Further details on optical engineering aspects are reported in that paper. 

\begin{figure}
  \begin{center}
    \includegraphics[width=0.48\textwidth]{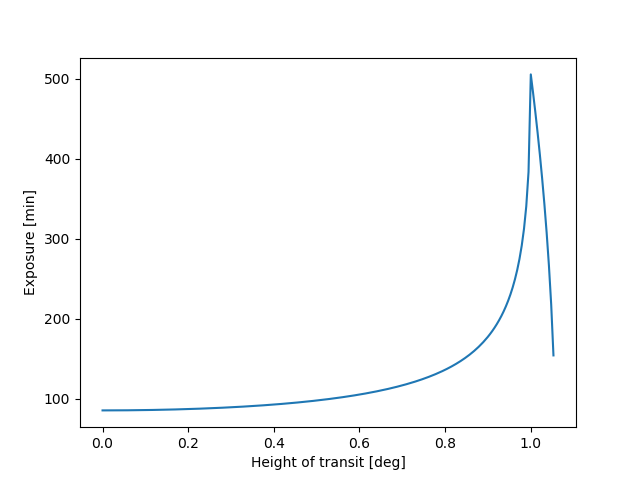}
    \includegraphics[width=0.48\textwidth]{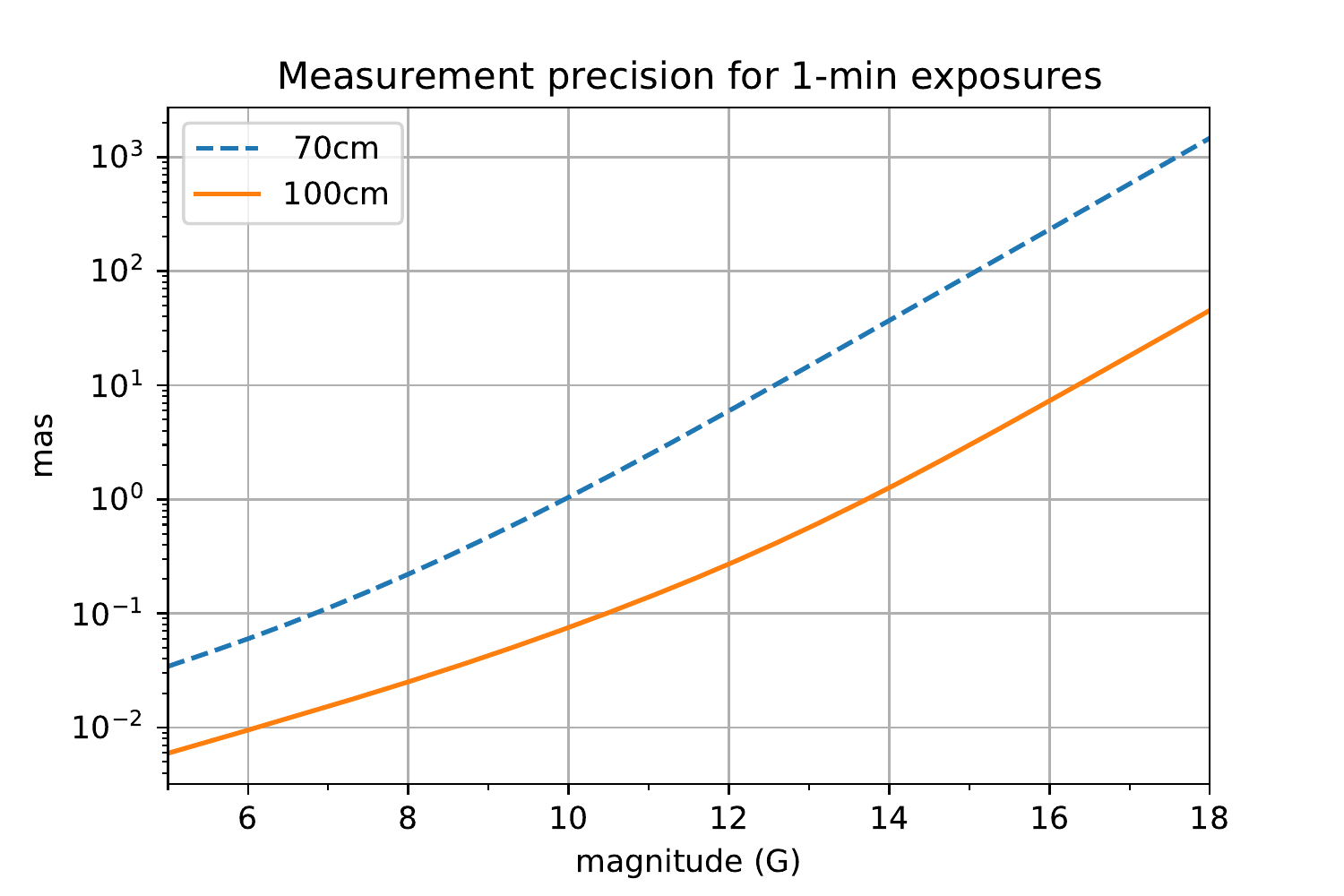}
    \caption{\label{fig:ExpTimeand1minAcc} Exposure time vs. Ecliptic latitude (left pane) and measurement precision for a 1-minute exposure (right pane).}
  \end{center}
\end{figure}

\subsection{Detection system}
\label{sec:detector}
The detection system considered for our performance evaluation is an assembly of 66 devices (CCD or sCMOS) with logical format $4k \times 4k$ and pixel size $4\,\mu m$, providing adequate PSF sampling ($\sim 50\,mas$) with the design effective focal length $EFL = 15\,m$. 
The sky area corresponding to one chip is $S_C \simeq 3'45''$. Notably, the detector exploits only a fraction (0.258 square deg) of the total $1.257$ square deg diffraction limited field; the provisional choice is of course subject to optimization. 
In particular, the selected pixel size is compatible with existing CMOS devices, but it requires some developments on CCDs.

\subsection{Operations}
\label{sec:operations}
At any time, an annular region with width $\sim 3'.75$, and mean radius of $1^\circ$, centred on the Sun, is observed. The sidereal motion associated to the Earth's orbit corresponds to a displacement of the observed region along the Ecliptic equator by $\sim 1^\circ$ per day, corresponding to a drift speed $V_D \simeq 0''.\!\!0411/s$. 

The observing mode shall be defined by a trade-off between step-and-stare and step-and-stare plus scanning; the latter option is feasible in case of usage of CCDs, as the satellite may be operated at a constant revolving rate around the optical axis, plus a tilt by $2''\!\!.5$ every minute, on average, to chase the Sun and retain it close to the instrument optical axis. A practical implementation may consist in application of a $\sim 14''$ displacement every five minutes, allowing for $\sim 30\,s$ dead time between observations for re-pointing and stabilisation. This period may be used to advantage by an on-board metrology system. 

The total exposure time, due to the superposition of the annular field throughout subsequent pointing positions, is not constant, but it changes with ecliptic latitude. On the ecliptic, the field is mostly orthogonal to the sidereal motion, and it is crossed in $S_C / V_D = 3'45'' / 0''\!\!.0411\,s \simeq 5474\,s$, i.e. $\sim 1.5$ hours. 
Close to the northern or southern edge, an object crosses a much larger section of the ring focal plane, reaching a total observing time of more than 8 hours. 
The dependence of observing time on ecliptic latitude is shown in Fig.\,\ref{fig:ExpTimeand1minAcc} (left).

Therefore, although the whole Ecliptic strip within $\pm 1^\circ$ is observed at the same time, the efficiency is not uniform. Repeated observation of the same stars over several chips is beneficial to calibration. 

\section{Expected Performance }
\label{sec:performance}  

\begin{table}
  \begin{center}
    \begin{tabular}{ccc}
        \hline
        \multirow{2}{*}{Time (yr)} & \multicolumn{2}{c}{$\sigma_\gamma\times10^7$} \\\cline{2-3}
                  & $D_1=70\,\mathit{cm}$ & $D_1=1\,\mathit{m}$ \\
        \hline\hline
        $1$ & $6.13$ & $0.841$ \\
        $2$ & $4.34$ & $0.595$ \\
        $3$ & $3.54$ & $0.485$ \\
        $4$ & $3.07$ & $0.420$ \\
        $5$ & $2.74$ & $0.376$ \\
        \hline
    \end{tabular}
    \caption{\label{tab:EstGammaAcc} Estimation of $\sigma_\gamma$ for the two configurations at 1-yr time steps.}
  \end{center}
\end{table}

We performed some simple simulations to have a quick estimation of the expected final mission accuracy.

First, we estimated the measurement performance $\sigma_{\psi1}$ for a 1-minute exposure within the $5\leq G\leq18$ magnitude range, and for the two different configurations described in Sect.~\ref{sec:TelOptDes}, namely for a primary mirror of diameter $D_1=0.70\,m$ and $D_1=1\,m$ respectively. The results are reported on the right in Fig.~\ref{fig:ExpTimeand1minAcc}.

Then we extracted all the sources in the same magnitude range and with ecliptic latitude $-1.5^\circ\leq\lambda\leq1^.5\circ$ from the Gaia DR2 catalog. This produced a total of $\sim11$~million objects.

We used Eq.~\eqref{eq:sigma_gamma_3} to evaluate the approximate individual $\sigma_\gamma$ for each selected source. Since the objects are observed on a ring of a $1^\circ$ radius and a thickness of about $0^\circ\!\!.2$, for our purposes it was sufficient to approximate all the deflections to the same value, that is $\delta\psi(1^\circ)\simeq466.713~\mathit{mas}$. On the other hand, the approximate $\sigma_\psi$ depended on the total exposure time, namely on the ecliptic latitude of the source. We thus considered
\begin{equation}
    \sigma_\psi(i)=\frac{\sigma_{\psi1}}{\sqrt{\Delta t_i}},
\label{eq:sigmapsi_sim}
\end{equation}
where $\Delta t(i)$ is the total exposure of the object $i$, in minutes, as derived from its ecliptic latitude and from the considerations of Sect.~\ref{sec:operations}. Substituting these quantities in Eq.~\eqref{eq:sigma_gamma_3} we obtained the accuracy on the PPN-$\gamma$ estimation from each object after one year of observations.

\begin{figure}
  \begin{center}
    \includegraphics[width=0.95\textwidth]{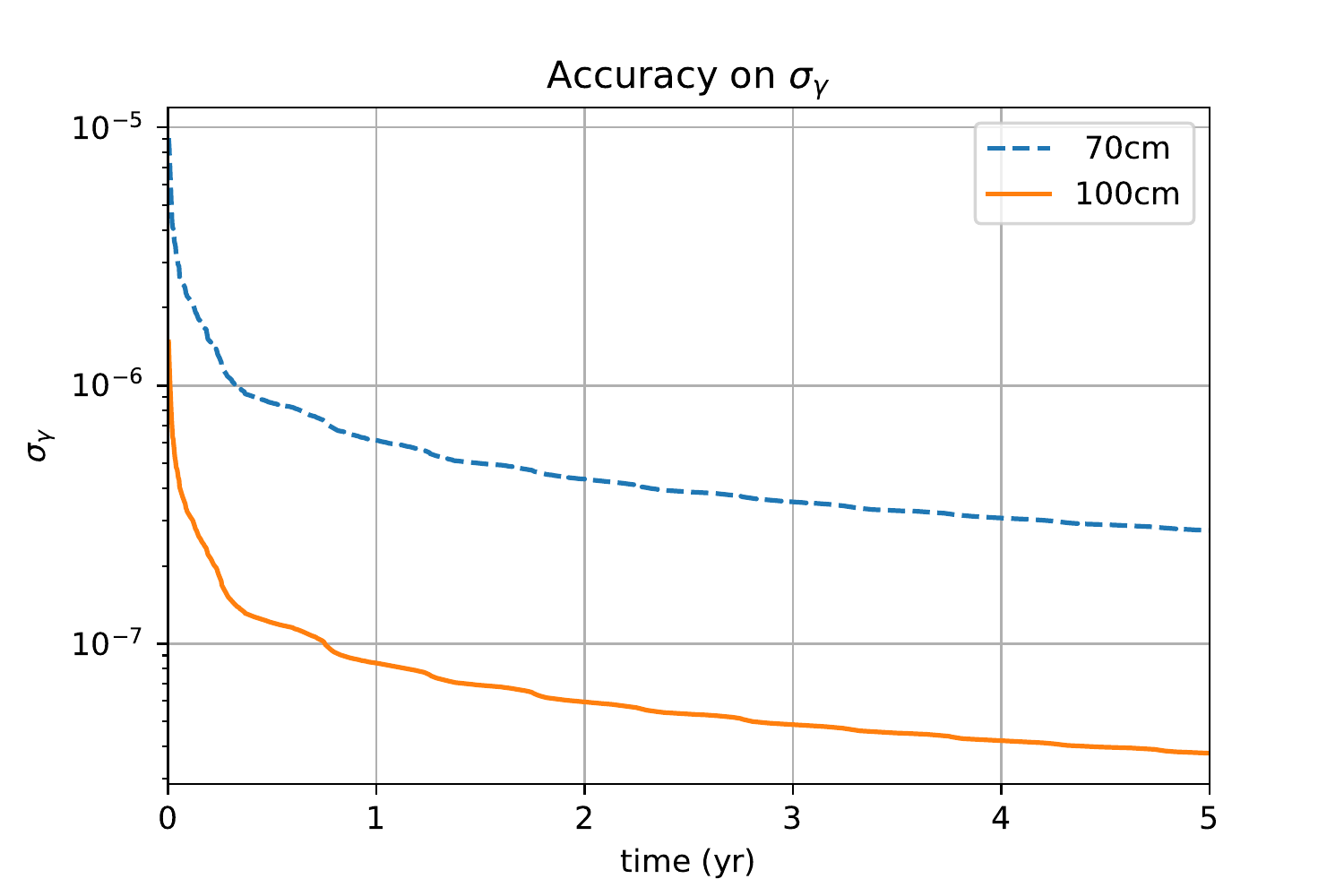}
    \caption{\label{fig:EstGammaAcc} Estimation of the accuracy on $\gamma$ as function of time for the two instrument configurations.}
  \end{center}
\end{figure}

Obviously, the final accuracy after one year of observations is represented by the error on the weighted mean on the $n$ sources
\begin{equation}
    \sigma_\psi(1yr)=\left(\sum_1^n\frac{1}{\sigma_\psi(i)^2}\right)^{-1},
\label{eq:sigmagamma_1yr}
\end{equation}
while the accuracy after $N$ years will be $\sigma_\psi(Nyr)=\sigma_\psi(1yr)/\sqrt{N}$. Finally, we obtained the plot of Fig.~\ref{fig:EstGammaAcc} by grouping the stars by $1^\circ$ bins of ecliptic longitude and making the approximation that all the contribution of each star to the total accuracy falls in a single bin. The final results are reported in Tab.~\ref{tab:EstGammaAcc}, where it can be observed that the configuration with the larger $D_1$ is suitable to reach an accuracy at the $10^{-8}$ level.

\section{Conclusions}
\label{sec:conclusions}  
We evaluated the performance of an implementation of the AGP mission concept in a configuration similar to the one considered for the RAFTER telescope.\cite{Riva2020RAFTER} The inverted occulter principle exploited in this optical design might be able to drastically reduce the diffraction produced by the supporting structure of the traditional occulters. On the other hand, an annular field of view is particularly suited for the purposes of this mission, namely to probing the effect of the curvature of space-time around the Sun on the light paths.

Preliminary evaluations based on a simplified but realistic description of the system suggest that a 70-cm diameter instrument is able to reach an accuracy of $\sim10^{-7}$ after one or more years. Similarly, a 1-meter diameter instrument is able to reach an accuracy of $\sim10^{-8}$, which candidates AGP as the best option to estimate the PPN-$\gamma$ parameter. Our considerations are based on a mission lifetime of up to 5 years, and although the order-of-magnitude performance can be reached already after 1 year, extending the mission duration for a longer period allows to consolidate our confidence on the results by strengthening the calibration. Furthermore, it makes up for a larger than 2 improvement factor on $\sigma_\gamma$ and opens the opportunity to allocate some observing time for other scientific goals, including targets of opportunity.

It has to be stressed that a $\sigma_\gamma\sim10^{-8}$ in a region where $\delta\psi\simeq5\cdot10^-1$~arcsec, from the approximate formula of Eq.~\eqref{eq:sigma_gamma_3}, is equivalent to realize a single measure with $\sigma_\psi\sim10^{-9}$~arcsec. As this level of accuracy is beyond the Post-Newtonian one, the consequences on the modelling requirements has thus to be further investigated, and they might require the extension of the astrometric model to the 2PN regime, i.e.\ up to the $(v/c)^4$ order.

The RAFTER telescope class, due to the circular symmetry of its optical response, is particularly suited to the AGP measurement of deflection, since the observable has circular symmetry with respect to the Sun, located close to the optical axis throughout observations. Also, since the instrument ensures decoupling between radial and azimuthal measurement, some classes of systematic errors are rejected to a  high degree. Moreover, as discussed in the related paper, the RAFTER configuration, thanks to exploitation of circular symmetry at any stage, is intrinsically robust with respect to astrometric performance preservation against optical system perturbations. It is also very compact, and therefore suited to convenient allocation on a 
spacecraft, also in combination with other instruments. 

The estimated AGP performance is quite appealing, and will be further improved by forthcoming optimization activity, both on the instrumental package and on the measurement modelling aspects. An obvious upgrade path consists in an increase of the detector size, since the current design exploits about 20\% of the diffraction limited field of view provided by the current optical system. The obvious impact on the mission costs is to be traded off against a performance improvement by a further factor 2. Other improvements may be achieved by obvious increases in the telescope diameter, increasing both collecting area and resolution, but also by reducing the radius of the annular field of view. This results in more stringent requirements on the coronagraphic system, but taking the measurements to a region featuring larger deflection (from Eq.\,\ref{eq:deflMisner}) increases their effectiveness (Eq.\,\ref{eq:sigma_gamma_3}), even with the same angular uncertainty on source position estimates. 

The AGP concept proves therefore to be flexible, robust and able to provide reliable and accurate measurements at the forefront of Fundamental Physics. 

\acknowledgments          

The INAF activity has been partially funded by a grant from the Italian Ministry of Foreign Affairs and International Cooperation, and by the Italian Space Agency (ASI) under contracts 2014-025-R.1.2015 and 2018-24-HH.0. 

\bibliographystyle{spiebib}
\bibliography{mybibl}

\begin{thebibliography}{10}

\bibitem{2019NCimR..42..443C}
M.~{Crosta}, ``{Astrometry in the 21st century. From Hipparchus to Einstein},''
  {\em Nuovo Cimento Rivista Serie}~{\bf 42}, pp.~443--510, Oct. 2019.

\bibitem{Will2006}
C.~M. Will, ``The confrontation between general relativity and experiment,''
  {\em Living Reviews in Relativity}~{\bf 9}(3), 2006.

\bibitem{Bertotti2003}
B.~{Bertotti}, L.~{Iess}, and P.~{Tortora}, ``{A test of general relativity
  using radio links with the Cassini spacecraft},'' {\em Nature (London)}~{\bf
  425}, pp.~374--376, Sept. 2003.

\bibitem{2003AAp...399..337V}
A.~{Vecchiato}, M.~G. {Lattanzi}, B.~{Bucciarelli}, M.~{Crosta}, F.~de~Felice,
  and M.~{Gai}, ``{Testing general relativity by micro-arcsecond global
  astrometry},'' {\em Astron. Astrophys.}~{\bf 399}, pp.~337--342, Feb. 2003.

\bibitem{2010IAUS..261..315H}
D.~{Hobbs}, B.~{Holl}, L.~{Lindegren}, F.~{Raison}, S.~{Klioner}, and
  A.~{Butkevich}, ``{Determining PPN {\ensuremath{\gamma}} with Gaia's
  astrometric core solution},'' in {\em Relativity in Fundamental Astronomy:
  Dynamics, Reference Frames, and Data Analysis},  S.~A. {Klioner}, P.~K.
  {Seidelmann}, and M.~H. {Soffel}, eds.,  {\bf 261}, pp.~315--319, Jan. 2010.

\bibitem{Butkevich2021inPrep}
A.~G. {Butkevich}, A.~{Vecchiato}, M.~G. {Lattanzi}, B.~{Bucciarelli},
  M.~{Gai}, and M.~{Crosta}, ``{The PPN-$\gamma$-parallax correlation in
  astrometry - II. Effects of a parallax bias on PPN-$\gamma$},'' {\em Astron.
  Astrophys. (in Preparation)} , 2021.

\bibitem{2009ExA....27...27T}
S.~G. {Turyshev}, M.~{Shao}, K.~L. {Nordtvedt}, H.~{Dittus}, C.~{Laemmerzahl},
  S.~{Theil}, C.~{Salomon}, S.~{Reynaud}, T.~{Damour}, U.~{Johann},
  P.~{Bouyer}, P.~{Touboul}, B.~{Foulon}, O.~{Bertolami}, and J.~{P{\'a}ramos},
  ``{Advancing fundamental physics with the Laser Astrometric Test of
  Relativity. The LATOR mission},'' {\em Experimental Astronomy}~{\bf 27},
  pp.~27--60, Dec. 2009.

\bibitem{Gai2020ASTRA}
M.~{Gai}, Z.~{Qi}, M.~{Lattanzi}, and {et al.}, ``{The ASTRA project: a doorway
  to future astrometry},'' {\em Proc.\ SPIE} {\bf 11451}, p.~114514I, Dec.
  2020.

\bibitem{Riva2020RAFTER}
A.~{Riva}, M.~{Gai}, A.~{Vecchiato}, D.~{Busonero}, , M.~G. {Lattanzi},
  F.~{Landini}, Z.~{Qi}, and Z.~{Tang}, ``{RAFTER: Ring Astrometric Field
  Telescopefor Exo-planets and Relativity},'' in {\em This Conference},  {\em
  SPIE Conference Series}, 2020.

\bibitem{Misner1973}
C.~W. {Misner}, K.~S. {Thorne}, and J.~A. {Wheeler}, {\em Gravitation}, San
  Francisco: W.H.~Freeman and Co., 1973.

\bibitem{AGP_Landini16}
F.~{Landini}, A.~{Riva}, M.~{Gai}, {\em et~al.}, ``{Stray light evaluation for
  the astrometric gravitation probe mission},'' in {\em Proc.\ SPIE},  {\em
  Society of Photo-Optical Instrumentation Engineers (SPIE) Conference Series}
  {\bf 9907}, p.~990741, Aug. 2016.

\bibitem{Antonucci_2020}
E.~{Antonucci}, M.~{Romoli}, V.~{Andretta}, and {et al.}, ``{Metis: the Solar
  Orbiter visible light and ultraviolet coronal imager},'' {\em Astron.
  Astrophys.}~{\bf 642}, p.~A10, 2020.

\bibitem{Fineschi_2020}
S.~{Fineschi}, G.~{Naletto}, M.~{Romoli}, and {et al.}, ``{Optical design of
  the multi-wavelength imaging coronagraph Metis for the solar orbiter
  mission},'' {\em Experimental Astronomy}~{\bf 49}(3), pp.~239--263, 2020.

\bibitem{Metis_Landini2016}
F.~{Landini}, M.~{Romoli}, G.~{Capobianco}, and {et al.}, ``{Improved stray
  light suppression performance for the solar orbiter/METIS inverted external
  occulter},'' in {\em Solar Physics and Space Weather Instrumentation V},
  S.~{Fineschi} and J.~{Fennelly}, eds., {\em Society of Photo-Optical
  Instrumentation Engineers (SPIE) Conference Series} {\bf 8862}, p.~886204,
  Sept. 2013.

\bibitem{Korsch77}
D.~{Korsch}, ``{Anastigmatic three-mirror telescope.},'' {\em Appl. Opt.}~{\bf
  16}, pp.~2074--2077, Aug. 1977.

\bibitem{Korsch80}
D.~{Korsch}, ``{Design and optimization technique for three-mirror
  telescopes},'' {\em Appl. Opt.}~{\bf 19}, pp.~3640--3645, Nov. 1980.

\end{thebibliography}

\end{document}